\documentclass[prb,twocolumn,showpacs,amsmath,amssymo]{revtex4}
\usepackage{graphicx}
\usepackage{dcolumn}
\usepackage{bm}

\begin{document}
\title[c]{Structural Basis for the Anomalously Low Spontaneous Polarisation
Values of the Polar Phase of Sr$_{1-x}$Ca$_{x}$TiO$_{3}$
(x=$0.02$, $0.04$): Evidence for a Ferrielectric Ordering    }
\author{Sanjay Kumar Mishra}%
\affiliation{%
Solid State Physics Division, Bhabha Atomic Research Center,
Trombay, Mumbai 400 085, India}
\author{Dhananjai Pandey}
\email{dpandey@bhu.ac.in, dpandey_bhu@yahoo.co.in}
\affiliation{School of Materials Science and Technology, Institute
of Technology Varanasi-221005, India}
\date{\today}

\begin{abstract}
Full pattern Le-Bail refinement using x-ray powder diffraction
profiles of Sr$_{1-x}$Ca$_{x}$TiO$_{3}$ for x=0.02, 0.04 in the
temperature range 12 to 300 K reveals anomalies in the unit cell
parameters at 170, 225 K due to an antiferrodistortive (cubic to
tetragonal \textit{I4/mcm}) phase transition and at $\sim 32, \sim
34 K$ due to a transition to a polar phase (tetragonal
\emph{I4/mcm} to orthorhombic\textit{ Ic2m}), respectively. The
lower transition temperatures obtained by us are in excellent
agreement with those reported on the basis of the dielectric
studies by Bednorz and Muller,\cite{jgb} who attributed these to
ferroelectric transition. Rietveld analysis of the diffraction
profiles of the polar phase reveals off-centre displacements of
both $Sr^{2+}/Ca^{2+}$ and Ti$^{4+}$ ions in the X-Y plane along
$\langle {110} \rangle$ pseudocubic directions, in agreement with
the experimentally reported direction of easy polarization by
Bednorz and Muller, but the resulting dipole moments are shown to
be ferrielectrically coupled in the neighbouring $(001)$ planes
along the $[001]$ direction leading to anomalously low values of
the spontaneous polarization at 12 K.
\end{abstract} 
\maketitle SrTiO$_{3}$ has been the paradigm of structural phase
transition studies.\cite{lg} It is also one of the most common
ingredients of BaTiO$_{3}$-based ceramic capacitors.\cite{rc} The
room temperature cubic perovskite structure of SrTiO$_{3}$ (space
group \textit{{$Pm \bar 3 m$}} ) transforms to a tetragonal phase
in the \textit{{I4/mcm}} space group below 105 K due to the
freezing of one of the triply degenerate R$_{25}$ modes in which
the adjacent octahedra rotate about one of the cubic $\langle
{100} \rangle$ directions in an anti-phase manner.\cite{gs}
Although the tetragonal phase is reported to be stable down to the
lowest temperature,\cite{kia} its dielectric constant gradually
increases on cooling and becomes nearly two orders of magnitude
greater at $\sim 4  K$ below which it levels off. While the
increasing value of the dielectric constant at low temperatures is
linked to a zone centre soft mode,\cite{lg} its saturation below 4
K has been attributed to quantum fluctuations (i.e., zero point
vibrations) of the lattice which stabilize the paraelectric
state.\cite{mull} $Ca^{2+}$ substitution not only increases the
antiferrodistortive phase transition temperature\cite{ub,skm}  but
also suppresses the quantum fluctuations, as revealed by the
appearance of a peak in the temperature variation of the
dielectric constant. In fact, the quantum fluctuations can be
suppressed by several ways, such as by the application of external
stress,\cite{hu} electric field\cite{paf} and suitable chemical
substitutions like Ca$^{2+}$, Ba$^{2+}$, Pb$^{2+}$ at the
Sr$^{2+}$ site and oxygen isotope (O$^{18}$) at the O$^{2-}$
sites.\cite{jgb,meg} It has been shown by Bednorz and
Muller\cite{jgb} that above a critical Ca$^{2+}$ concentration of
x$_{c}$= 0.0018, XY type 'ferroelectricity' is stabilized in
SrTiO$_{3}$.The ferroelectric transition temperature $(T_{c})$
rises with Ca$^{2+}$ concentration (x) upto $x = 0.016$ according
to the relationship $T_{c} \sim (x-x_{c})^{0.5}$ as expected for
quantum transitions.\cite{jgb} In the composition range
$0.016<x<0.12$, the transition temperature is nearly composition
independent $(T_{c}\sim 35 K)$ and the variation of dielectric
constant with temperature is increasingly more smeared out as the
Ca$^{2+}$ content increases. This smearing out of the transition
has been attributed by Bednorz and Muller\cite{jgb} to the
formation of mesoscopic 'ferroelectric' domains caused by the
random fields induced by some of the Ca$^{2+}$ occupying Ti$^{4+}$
sites, thereby forming $Ca^{2+}- V_{o}(V_{o}$ is oxygen vacancy)
dipoles. Ranjan et al\cite{rr} have argued that this smearing
could as well be the result of a frustration caused by the
competing ferroelectric and antiferroelectric instabilities for
these compositions, since the nature of the phase transition in
$Sr_{1-x}Ca_{x}TiO_{3}$ (SCT) changes from 'ferroelectric` for low
$Ca^{2+}$ doping levels to antiferroelectric for high $Ca^{2+}$
concentrations with $x>0.12$. Raman scattering studies\cite{ro}
reveal the presence of both the antiferroelectric and
ferroelectric modes for x=0.12 in support of the frustration
model.

For $Ca^{2+}$ concentrations ${x_{c}<x<0.12}$, the dielectric
constant of SCT in the 'ferroelectric' state decreases with
increasing x but its value $(>10^{4})$ remains an order of
magnitude larger than that of the conventional ferroelectrics like
$BaTiO_{3}$.\cite{jgb} On the otherhand, the measured saturation
polarization values $(\sim 0.5 \mu C/cm^{2})$ are anomalously
low\cite{ubz,tm} in the 'ferroelectric' phase as compared to that
in BaTiO$_{3}$, which is $\sim 25 \mu C/cm^{2}$. There is no
satisfactory explanation for such a low value of saturation
polarization $(\sim 0.5 \mu C/cm^{2})$in a 'ferroelectric' phase
with extremely high dielectric constant. The present work seeks to
provide a structural basis for this low value of spontaneous
polarization.

No attempt has so far been made to examine the structure of the
'ferroelectric' phase of SCT for $x_{c}> 0.0018$ using diffraction
techniques. In the present work, we have investigated the phase
transitions in SCT for x=0.02 and 0.04 (SCT02 and SCT04) in the
temperature range 12 to 300 K by Le-Bail and Rietveld techniques
of x-ray  powder diffraction, with a view to capture the
structural signatures of the antiferrodistortive and
'ferroelectric' transitions. Our results suggest that the
anomalously low spontaneous polarization value of SCT is linked
with the 'ferrielectric' nature of the hitherto regarded
'XY-ferroelectric' phase of SCT. The calculated spontaneous
polarization values, obtained using the Rietveld refined atomic
positions and Born-effective charges, as given in Ref.\cite{phg},
are found to be in excellent agreement with the values reported by
Bianchi et al\cite{ubz} and Mitsui and Westphal.\cite{tm}

SCT02 and SCT04 powders were prepared by solid-state
thermochemical reaction in the appropriate stoichiometric mixtures
of $SrCO_{3}, CaCO_{3}$ and $TiO_{2}$ at 1423 K for 6 h in alumina
crucibles. The calcined powders were sintered at 1573 K for 6 h.
The sintered pellets were crushed to fine powder and subsequently
annealed at 823 K for 12 h for removing the strain induced, if
any, during the crushing process before using them for diffraction
studies. Temperature dependent powder x-ray diffraction studies in
the 12 to 300 K range were carried out using a 18 kW Rigaku (RINT
2000/ PC series) rotating anode based high resolution x-ray powder
diffractometer operating in the Bragg-Brentano focusing geometry.
The diffractometer is fitted with a He- closed cycle refrigerator
based low temperature attachment and also a curved crystal
monochromator in the diffraction beam. Data were collected in the
continuous scan mode at a scan speed of 1 degree per minute and
step interval of 0.01 degree. The package 'Fullprof'\cite{jr}was
used for Le-Bail and Rietveld analysis of the x-ray diffraction
data.

For pure SrTiO$_{3}$, the cubic to tetragonal antiferrodistortive
phase transition is accompanied with the doubling of the
elementary perovskite cell leading to characteristic splitting of
main perovskite peaks along with the appearance of superlattice
reflections.\cite{kia} The relationship between the tetragonal
cell parameters $(a_{t}, b_{t}, c_{t})$ with the elementary
perovskite cell parameters $(a_{p}, b_{p}, c_{p})$ is as follows:
$a_{t}= b_{t} \approx \surd 2 a_{p}, c_{t} \approx 2c_{p}$. The
Miller indices of the main perovskite and superlattice reflections
of the tetragonal phase with $a^{0}a^{0}c^{-}$ tilt system are all
even (eee) and all odd (ooo) integers, respectively, when indexed
with respect to a pseudocubic $2a_{p}\times 2b_{p}\times 2c_{p}$
unit cell (see Glazer\cite{glazer} for details). Figs. 1 (a) and
(b) depict the evolution of the 311 superlattice and 800
perovskite reflections as a function of temperature for SCT02 and
SCT04, respectively. It is evident from these figures that the 800
profile is a doublet at low temperatures. It is also evident from
this figure that the intensity of the superlattice reflection, as
also the splitting of the 008 peak, disappears at high
temperatures, as expected for the high temperature phase.
\begin{figure}
\includegraphics[height= 8.50 cm,width= 6.50 cm]{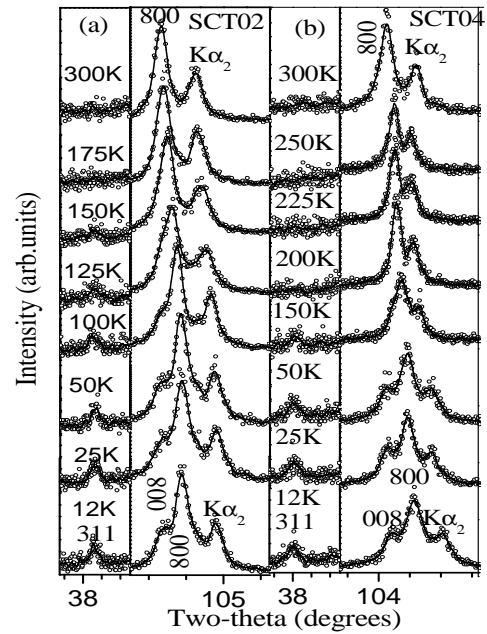}
\caption{\label{fig1}Evolution of the XRD profile of the 311
superlattice and $800$ perovskite reflections of (a)
$Sr_{0.98}Ca_{0.02}TiO_{3}(SCT02)$ and (b)
$Sr_{0.96}Ca_{0.04}TiO_{3}(SCT04)$. The indices are with respect
to a doubled pseudocubic perovskite cell. }
\end{figure}
In order to determine the phase transition temperatures, we
analysed the powder diffraction data in the $2\theta$ range of 20
to $110 \deg$ at various temperatures by Le-Bail
technique$^{17}$using the \textit{I4/mcm} space group. The
variations of the equivalent elementary perovskite cell parameters
with temperature, as obtained from the refined cell parameters of
the tetragonal phase, are plotted in Fig. 2. It is evident from
this figure that the cubic to tetragonal transition occurs around
$T_c= 170$ and 225 K for x=0.02 and 0.04, respectively. Combining
these two transition temperatures with the previously reported
values for $x= 0.0,\cite{ka} {0.007},\cite{ub} {0.01},{0.06}$ and
${0.12}\cite{skm}$ one gets a $dT_{c}/dx$ as $34.83 K/mole$ for $x
< 0.06$ and $ {12.54  K/mole}$ for $x>0.06$ showing a kink
corresponding to a change of slope $(dT_{c}/dx)$ at $x \sim 0.06$
in the $T_{c}$ versus 'x' plot, as noted by Mishra et al
earlier\cite{skm} but missed in a recent report.\cite{carp}
\begin{figure}
\includegraphics[height= 6.50 cm,width= 5.50 cm]{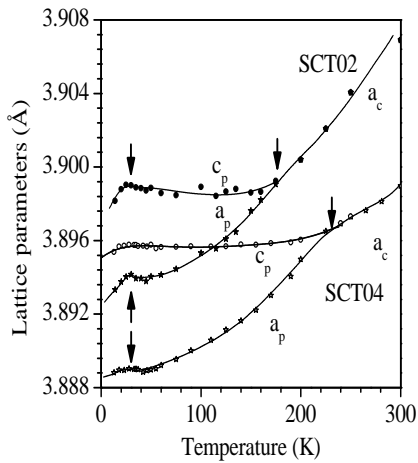}
\caption{\label{fig2}Evolution of the equivalent elementary
perovskite cell parameters of $Sr_{0.98}Ca_{0.02}TiO_{3}(SCT02)$
and $Sr_{0.96}Ca_{0.04}TiO_{3}(SCT04)$ as a function of
temperature showing the phase transition to the cubic phase around
170K and 225 K, marked with arrows. The anomaly $\sim 32$ K,
linked with the transition to the polar phase, is marked with
arrows. }
\end{figure}

Although the XRD profiles shown in Fig. 1 do not depict any
qualitative change in the 'ferroelectric phase' (i.e., below 35
K), the refined tetragonal cell parameters show anomalies around
30 K in Fig.2. Since the transition temperatures reported in the
literature for SCT02 and SCT04 are around $\sim 32  K$,\cite{jgb}
the anomalies in the cell parameters at $\sim 30  K$ are evidently
linked with the phase transition from the tetragonal paraelectric
phase to the lower temperature 'ferroelectric' phase. The
centrosymmetric \textit{I4/mcm }space group of the paraelectric
tetragonal phase cannot represent the structure of the
ferroelectric phase. The onset of ferroelectric polarization
transforms under \textit{$Pm \bar 3 m$} as the components of the
$\Gamma_{4}^{-}$ irreducible representation (IR). Anti-phase
tilting of oxygen octahedra transforms under \textit{$Pm \bar 3
m$} as the components of the $R_{4}^{+}$ IR.\cite{ht} Using the
software package "ISOTROPY 2000"\cite{ht} to couple
$\Gamma_{4}^{-}$ and $R_{4}^{+}$  and IRs, we obtained many space
groups. Since the paraelectric phase of SCT is tetragonal
\textit{(I4/mcm)}, we impose the constraint that the
'ferroelectric space' group is a subgroup of \textit{I4/mcm}. This
constraint reduced the number of possible space groups for the
ferroelectric phase to two, namely: \textit{I4cm} and
\textit{Ic2m} (bca setting of \textit{Ima2}). The Raman scattering
studies on SCT with $x=0.007$\cite{ub} have shown that the $E_{g}$
mode is split into two lines in the 'ferroelectric phase' and
hence the tetragonal \textit{I4cm} space group, for which $E_{g}$
mode should have remained a singlet, cannot represent the
ferroelectric phase. Thus, the orthorhombic space group
\textit{Ic2m} is the mostly likely space group of the
ferroelectric (polar) phase of SCT for which $E_{g}$ mode will
indeed be a doublet. The equivalent elementary perovskite cell
parameters obtained at various temperatures by Le-Bail analysis of
the full diffraction profiles in the $2\theta$ range 20 to 120
degrees using the \textit{Ic2m} space group are depicted in Figs.
3(a) and (b) for SCT02 and SCT04, respectively. The cell
parameters $(A_{o}, B_{o}, C_{o})$ of the orthorhombic
\textit{Ic2m} phase are related to the elementary perovskite cell
parameters $(a_{p}, b_{p}, c_{p})$ as: $A_{o} \approx \surd 2
a_{p}, B_{o}\approx \surd 2 b_{p}$  and $C_{o}= 2c_{p}$. It is
evident from this figure that the $a_{p}$ lattice parameter of the
tetragonal paraelectric phase splits below $\sim 32 K$ into
$a_{p}$  and  $b_{p}$ of the orthorhombic phase, in good agreement
with the transition temperature $(T_{c}\sim 32 K)$ at which the
dielectric peak is observed.\cite{jgb}
\begin{figure}
\includegraphics[height= 6.50 cm,width= 5.50 cm]{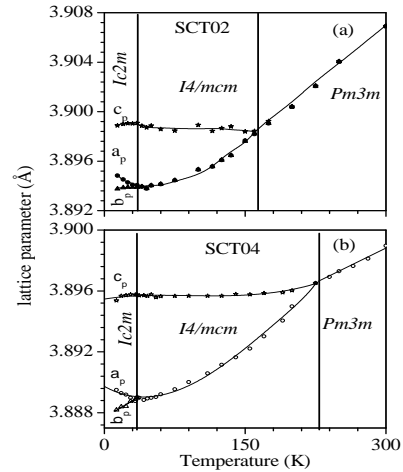}
\caption{\label{fig3}Evolution of the equivalent elementary
perovskite cell parameters of $Sr_{0.98}Ca_{0.02}TiO_{3} (SCT02)$
and $Sr_{0.96}Ca_{0.04}TiO_{3}(SCT04)$ as a function of
temperature showing the showing orthorhombic splitting of the a
parameter of the tetragonal phase below $\sim 32 K.$ }
\end{figure}

Having identified the most plausible space group for the polar
phase, we carried out Rietveld analysis of the 12 K XRD data in
the $2 \theta$  range 20 to 110 degree for SCT02 and SCT04 using
the \textit{Ic2m} (bca setting of \textit{Ima2}) space group. In
this space group, Sr/ Ca occupy the 4{b} Wyckoff site at ${(0 \pm
u, 0 \pm v, 0.25)}$ and the Ti occupies 4{a} Wyckoff site at ${(0,
0.5 \pm v, 0.5)}$. There are two types of oxygen atoms, O( 1) at
4{b} Wyckoff site at ${( 0 \pm u, 0.5\pm v , 0.5)}$ and O{(2)} at
8{c} site at ${(0.25 \pm u , 0.75 \pm v, 0 \pm w)}$. Pseudo-Voigt
function was used to define the profile shapes while background
was estimated by linear interpolation between fixed values. Except
for the occupancy parameters of atoms, which were fixed at their
nominal compositions, all other parameters were refined.Table 1
depicts the refined structural parameters for the ferroelectric
phase of SCT02 and SCT04, respectively.
\begin{table*}
\caption{\label{tab1}Refined structural parameters of the
ferrielectric phase of SCT02 and SCT04 at 12 K using orthorhombic
space group \textit{Ic2m}.}
\begin{ruledtabular}
\begin{tabular}{ccccccccc}
Atoms&\multicolumn{3}{c}{$Sr_{0.98}Ca_{0.02}TO_{3}(SCT02)$}&\multicolumn{5}{c}{$Sr_{0.98}Ca_{0.04}TO_{3}(SCT04)$}\\
 \hspace{0mm} &\multicolumn{1}{c} x &\multicolumn{1}{c} y &\multicolumn{1}{c} z &\multicolumn{1}{c} B $({\rm \AA})^{2}$ &\multicolumn{1}{c} x &\multicolumn{1}{c} y &\multicolumn{1}{c} z&\multicolumn{1}{c} B$({\rm \AA})^{2}$\\
 \hline
 \hspace{0mm}Sr/Ca &\multicolumn{1}{c} {0.0018(9)} &\multicolumn{1}{c} {0.0091(7)} &\multicolumn{1}{c} {0.25} &\multicolumn{1}{c} {0.67(7)} &\multicolumn{1}{c} {0.0028(2)}&\multicolumn{1}{c} {0.0083(5)}&\multicolumn{1}{c} {0.25} &\multicolumn{1}{c} {0.89(5)}\\
 \hspace{0mm}Ti &\multicolumn{1}{c} {0.00} &\multicolumn{1}{c} {0.4894(7)} &\multicolumn{1}{c} {0.50} &\multicolumn{1}{c} {0.008(5)} &\multicolumn{1}{c} {0.00}&\multicolumn{1}{c} {0.4891(9)} &\multicolumn{1}{c} {0.50} &\multicolumn{1}{c} {0.011(7)}\\
 \hspace{0mm}O1 &\multicolumn{1}{c} {-0.023(3)} &\multicolumn{1}{c} {0.552(7)} &\multicolumn{1}{c} {0.25} &\multicolumn{1}{c} {0.052(1)} &\multicolumn{1}{c} {-0.021(6)}  &\multicolumn{1}{c} {0.5598(2)} &\multicolumn{1}{c} {0.25} &\multicolumn{1}{c} {0.24(7)}\\
 \hspace{0mm}O2 &\multicolumn{1}{c} {0.267(3)} &\multicolumn{1}{c} {0.7148(7)} &\multicolumn{1}{c} {0.001(2)} &\multicolumn{1}{c} {0.67(7)} &\multicolumn{1}{c} {0.2588(2)} &\multicolumn{1}{c} {7101(7)} &\multicolumn{1}{c} {0.003(4)} &\multicolumn{1}{c} {0.70(7)}\\
 \hline
 \hspace{10mm} &\multicolumn{8}{c}  {Lattice Parameters ($\rm \AA $)}\\
 \hspace{0mm} &\multicolumn{1}{c} {A=5.5073(3)} &\multicolumn{1}{c} {B=5.5066(2)} &\multicolumn{1}{c} {C=7.7977(1)} &\multicolumn{3}{c} {A=5.4979(3)} &\multicolumn{1}{c} {B=5.4997(2)} &\multicolumn{1}{c} {C=7.7908(1)}\\
 \hline
 \hspace{0mm} &\multicolumn{1}{c} {R$_{p}$=11.90} &\multicolumn{1}{c} {R$_{wp}$=18.73} &\multicolumn{1}{c} {R$_{e}$=12.97} &\multicolumn{1}{c} {$\chi^{2}= 2.08$}&\multicolumn{1}{c} {R$_{p}$=8.87} &\multicolumn{1}{c} {R$_{wp}$=13.0} &\multicolumn{1}{c} {R$_{e}$=9.36} &\multicolumn{1}{c} {$\chi^{2}= 1.94$}\\
 \end{tabular}
 \end{ruledtabular}
 \end{table*}

Fig. 4 depicts schematically the ionic displacements in the
orthorhombic ferroelectric phase of SCT02/04 with respect to the
ideal positions in the paraelectric phase. The crystal structure
of ferroelectric phase consists of the stacking of TiO, Sr/CaO,TiO
and Sr/CaO layers at $z=0$, ${1/4}$ ,${1/2}$ and ${3/4}$ of the
unit cell of ferroelectric phase, as shown in Fig. 4. All the ions
in these layers are displaced with respect to their ideal
positions in cubic perovskite structure. In the layers at z=
${1/4}$ and ${3/4}$, the Sr$^{2+}$/Ca$^{2+}$ and O$^{2-}$ ions
have component displacements both along $\pm [100]$ and $[010]$
directions of the orthorhombic cell. However, the component
displacements along $\pm [100]$ in the z=$1/4$ and z=$3/4$ planes
cancel out leaving net effective displacement only along the
$[010]$ direction of the orthorhombic phase. Similarly, for the
layers at $z={0}$ and ${1/2}$, the net displacements of Ti$^{4+}$
and O$^{2-}$ ions are along the $[0\bar{1} 0]$ direction of
orthorhombic ferroelectric phase. Thus the direction of
polarization in the TiO as well as Sr/CaO layers is along $\pm
[010]$ direction of the orthorhombic phase. This is in agreement
with the observation of Bednorz and Muller\cite{jgb} that the
$<110>$ pseudo-cubic direction, which becomes $<010>$ in the
orthorhombic cell, is the easy direction of polarization. However,
results of our refinements reveal ferrielectric coupling of the
dipole moments residing in the neighbouring $(001)$ layers, in
contrast to the suggestions of Bednorz and Muller\cite{jgb} for
ferroelectric order, since the dipole moments are along the $[0
\bar{1} 0]$  and $[010]$ directions in the alternate $(001)$
layers stacked in the [001] direction. Using the Born effective
charges given in Ref \cite{phg} for $Sr/Ca (q= 2.56 C), Ti (q=7.26
C), O (-3.2733 C)$ alongwith the refined positional coordinates
given in Table 1, we obtain a spontaneous polarization of $0.98
\mu C/cm^{2}$ and $1.17 \mu C/cm^{2}$ for x= 0.02 and 0.04,
respectively. These values are in excellent agreement with the
values reported by Bianchi et al\cite{ubz} and Mitsui and
Westphal\cite{tm} on the basis of hysteresis loop measurements.
Thus the extremely low value of spontaneous polarization, in spite
of the very high dielectric constant, of SCT02 and SCT04 is due to
the partial cancellation of the dipole moments as a result of the
ferroelectric coupling of the dipoles in the [001] direction of
the orthorhombic phase.
\begin{figure}
\includegraphics[height= 5.50 cm,width= 6.50 cm]{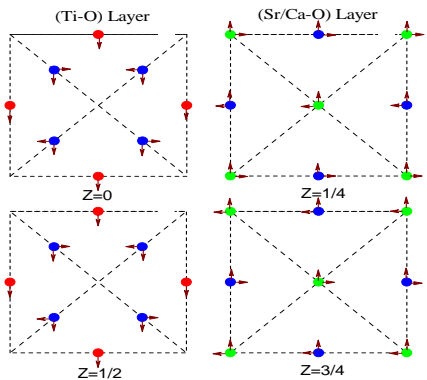}
\caption{\label{fig4}(color online) Schematic diagram showing the
ionic displacements in the orthorhombic ferroelectric phase of
Sr$_{0.98}$Ca$_{0.02}$TiO$_{3}$(SCT02) as layer wise and net
displacements of ions from its ideal position respectively.
Sr$^{2+}$/Ca$^{2+}$, Ti$^{4+}$ and O$^{2-}$ ions are represented
by filled circle with green, red and blue color, respectively. The
arrows indicate only the directions of displacement component
along the unit cell axes and not the magnitudes. }
\end{figure}

To summaries, we have shown that the orthorhombic distortion of
the lattice of Sr$_{1-x}$Ca$_{x}$TiO$_{3}$ (x=0.02, 0.04) below 35
K is clearly revealed through the Le-Bail analysis of the powder
x-ray diffraction profiles. Using symmetry arguments in
conjunction with Raman scattering results from literature, we have
proposed the \textit{Ic2m} space group for the low temperature
polar phase. This phase is not ferroelectric in nature, as
hitherto believed, but ferrielectric. The dipole moments in the
neighbouring (001) planes are coupled ferrielectrically in  the
[001] direction The value of spontaneous polarization calculated
using the Born-effective charges and the refined positional
coordinates obtained by Rietveld technique are in excellent
agreement with experimental values reported in the literature.

\end{document}